\documentclass[12pt]{iopart}

\usepackage{color}
\usepackage{graphicx}

\begin{document}

\title{Non-classical light from superconducting resonators coupled to voltage-biased
Josephson junctions}

\author{Bj\"orn Kubala, Vera Gramich, and Joachim Ankerhold}

\address{Institut f\"ur Theoretische Physik, Universit\"at Ulm, Albert-Einstein-Allee 11, 89069 Ulm, Germany\\
Center for Integrated Quantum Science and Technology, Albert-Einstein-Allee 11, 89069 Ulm, Germany }
\ead{bjoern.kubala@uni-ulm.de}

\begin{abstract}
 The interplay of the tunneling transfer of charges and the emission and absorption of light can be investigated in a set-up, where a voltage-biased Josephson junction is placed in series to a microwave cavity. Here, we concentrate on the emission process and show that due to the inherent nonlinearity of the Josephson junction tunneling Cooper-pairs can create a variety of non-classical states of light. Depending on experimental parameters and tuning the device can be described by effective Hamiltonians, indicating specific photon creation mechanisms which lead to strongly bunched or anti-bunched light emission. We derive explicit analytical expressions for the photon correlation function $g^{(2)}(\tau=0)$ for these different processes and investigate their full crossover numerically. Photon distribution functions show squeezing and other interesting non-Poissonian behavior.

\end{abstract}


%
%
%
%
%

\maketitle

\section{Introduction}
Cavity quantum electrodynamics (cQED) deals with the interaction of light quanta with atoms in optical cavities. This has led to an unprecedented level of control over quantum states and opened the door to the observation of a broad range of fundamental quantum phenomena. More recently, in an approach dubbed circuit-QED, atoms have been replaced by `artificial atoms', particularly designed (often superconducting) two-level systems, in microwave cavities.
The theory of quantum electrodynamics, however, has a much broader range and implies the interaction of fermionic (electrons) with bosonic (photons) matter in general.
In the non-relativistic regime, one of the most abundant processes in solid state physics is the transfer of charges, e.g.\ single electrons, quasi-particles or Cooper pairs, due to external voltage sources. In this field of quantum electronics  fascinating progress has been achieved as well in the last decades in fabricating devices on ever growing scales, with accurate control down to the level of individual charge carriers,  and in revealing a wealth of complex many-body phenomena.

Activities to combine these two previously basically distinct fields, cQED and quantum electronics,  have appeared only very recently, both in experiments and theory. One class of systems are semi-conductor or carbon-nanotube double quantum dots capacitatively coupled to
 a 1d-resonator and operated in the Coulomb blockade regime \cite{Frey:2012,Basset:2013, Delbecq:2013,Liu:2014}. Sequential tunneling of electrons is induced by a voltage bias and can be controlled by respective gate voltages. Theoretical descriptions \cite{Bergenfeldt:2013, Contreras:2013,Lambert:2013} reveal that the dot-cavity interaction mediated by virtual photons can lead to correlations in the currents through the individual dots and even to entanglement between the dots. Somewhat related set-ups consist of a suspended carbon-nanotube functioning as a mechanical resonator, where the nanomechanical motion couples to the charge transfer across the nanotube \cite{Steele:2009}. However, in both scenarios charge flow between the leads is always incoherent meaning that the coupling between the charge transfer units and the resonator is typically weak. Accordingly, these devices are operated close to the limit where the photon occupation in the resonator is zero.

This is in contrast to devices based completely on superconducting elements such as Josephson junction (JJs) or Cooper pair transistors embedded in 1d-resonators or transmission lines. They have been proposed theoretically as versatile experimental set-ups to produce non-classical photon states \cite{Gramich:2013} and to analyze the quantum-classical crossover in systems far from equilibrium \cite{Blencowe:2012}. Experimental realizations have been put forward, but so far have been limited to the regime of either very low \cite{Hofheinz:2011} or very high photon occupation \cite{Armour:2013, Chen:2013}. The rich physics behind these set-ups and the broad potential they may have in the future, has not been explored yet and awaits for combined experimental and theoretical efforts \cite{Padurariu:2012, Leppakangas:2013,Leppakangas:2014}.

In this work, we consider a set-up, where a voltage-biased JJ is placed in series to a cavity according to  \cite{Hofheinz:2011}, and focus on the photonic states of light in the cavity. The non-classical properties of these states are reflected in the microwave radiation leaking from the cavity, and are observable by absorption (or transmission and reflection) measurements. Explicit analytical results obtained in the low driving limit (low Josephson energy), where Cooper pair transfer occurs sequentially (Coulomb blockade regime) are compared to numerical calculations based on descriptions making use of rotating wave approximations (RWA) or treating the full problem. It turns out that this device creates a variety of non-classical states of light including strong anti-bunching and non-Poissionian photon distributions.

\section{System and steady state dynamics}

We start from a set-up, where a JJ with Josephson energy $E_J$ is subject to an external voltage bias $V$ and placed in series  to a resonator with a single mode frequency $\omega_0$. As shown in \cite{Gramich:2013}, this circuit can be described by
\begin{equation}\label{exact}
H=\hbar\omega_0 \, n -\frac{E_J}{2}\left( {\rm e}^{-i\eta}\, {\rm e}^{i\phi} \, {\rm e}^{i\omega_J t}+ h.c.\right)
\end{equation}
with occupation number $n=a^\dagger a$ and phase $\phi=\sqrt{\kappa} (a^\dagger +a)$ of the resonator with ground state width $\kappa=\hbar/2m\omega_0$, and conventional creation/annihilation operators $[a,a^\dagger]=1$. Further, $\omega_J=2 e V/\hbar$ denotes the driving frequency due to the voltage bias. Physically, the Josephson term captures the  transfer of Cooper pairs with the simultaneous exchange of resonator quanta: the phase $\eta$ is conjugate to the Cooper pair number operator $N$, i.e.\ $[\eta, N]=i$, so that ${\rm e}^{-i\eta}$ is a translational operator  in charge space, while  ${\rm e}^{i\phi}$ is the translational operator in the cavities' momentum space.

We first concentrate on the one-photon resonance $\omega_J\approx \omega_0$ and formulate (\ref{exact}) in a moving frame. The corresponding RWA then leads to
\begin{equation}\label{photon1}
H_{1}=\hbar\Delta \, n + i\frac{E_J^*}{2} \left[ : \left({\rm e}^{i\eta} a^\dagger -{\rm e}^{-i\eta} a\right)\, \frac{J_1(2\sqrt{\kappa n})}{\sqrt{n}}:\right]
\end{equation}
with de-tuning $\Delta=\omega_0-\omega_J$. Here,  $E_J^*=E_J {\rm e}^{-\kappa/2}$ denotes a renormalized Josephson coupling and we also introduced normal ordering to arrive at a compact notation in terms of Bessel functions. The latter ones collect creation/annihilation operators up to infinite orders, where particularly linear contributions are known to create coherent states and higher orders create squeezing.

According to the experimental situation, in the low temperature regime photon leakage out of the resonator into a heat bath of bosonic modes is the dominant source of decoherence. We describe the corresponding dynamics of the reduced density operator in the simple form of a Lindblad-type master equation, i.e.,
\begin{equation}\label{master1}
\frac{d\rho}{dt}= -\frac{i}{\hbar}[H_1, \rho] + {\cal L}_\gamma[\rho]
\end{equation}
with the dissipator ${\cal L}_\gamma[\rho]=(\gamma/2) (2 a\rho a^\dagger - n\rho-\rho n)$ and damping rate $\gamma$ being related to the cavity $Q$-factor via $Q=\omega_0/\gamma$. Extensions including low frequency voltage noise at the JJ are possible (see e.g.\ \cite{Gramich:2013}), but are  of no relevance here.
While in the long time limit the density operator does not approach a steady state, $d\rho_{\rm st}/dt\neq0$, due to the steadily increasing number of transferred Cooper pairs, observables like the Cooper pair current and the mean cavity occupation do reach stationary values, on which we focus in the remainder of this paper.
This way, one arrives for the cavity occupation at
\begin{eqnarray}
\langle n\rangle &=& \frac{{E}^*_J}{2\hbar\gamma} \langle :\left({\rm e}^{i\eta} a^\dagger +{\rm e}^{-i\eta} a\right) \frac{J_1(2\sqrt{\kappa n})}{\sqrt{n}}:\rangle\nonumber\\
\langle n^2\rangle &=& \langle n\rangle +\frac{{E}^*_J}{2\hbar\gamma}\langle :\left({\rm e}^{i\eta} a^\dagger +{\rm e}^{-i\eta} a\right) \left[ \sqrt{n} J_1(2\sqrt{\kappa n})-\sqrt{\kappa} J_2(2\sqrt{\kappa n})\right]:\rangle\, .
\end{eqnarray}
As a consequence, the photon correlation $g^{(2)}(0)=(\langle n^2\rangle -\langle n\rangle)/\langle n\rangle^2$, which indicates photon (anti)bunching on short time scales in the emitted radiation,
 takes the form
\begin{equation}
g^{(2)}(0)=\frac{1}{\langle n\rangle^2} \frac{{E}^*_J}{2\hbar\gamma}\langle :\left({\rm e}^{i\eta} a^\dagger +{\rm e}^{-i\eta} a\right) \left[ \sqrt{n} J_1(2\sqrt{\kappa n})-\sqrt{\kappa} J_2(2\sqrt{\kappa n})\right]:\rangle\, .
\end{equation}
In general, explicit results must be obtained numerically from the master equation (\ref{master1}). However, analytical progress can be made for low photon occupancy.

\section{Low photon occupancy}

In the regime $\kappa\langle n\rangle\ll 1$, the photon occupation in the cavity is small. Seen from the JJ, this corresponds to a domain, where Cooper pairs are transferred sequentially in presence of an environment (cavity) which prior to each transfer is in or close to thermal equilibrium (ground state at $T=0$). This regime is known also as Coulomb blockade (CB) regime, in contrast to the classical Josephson regime associated with a coherent flow of Cooper pairs. Hence,  JJ and cavity are only weakly coupled in the CB domain with basically no back-action from the cavity onto the junction.
\begin{figure}
\centering
\includegraphics[width=0.5\linewidth]{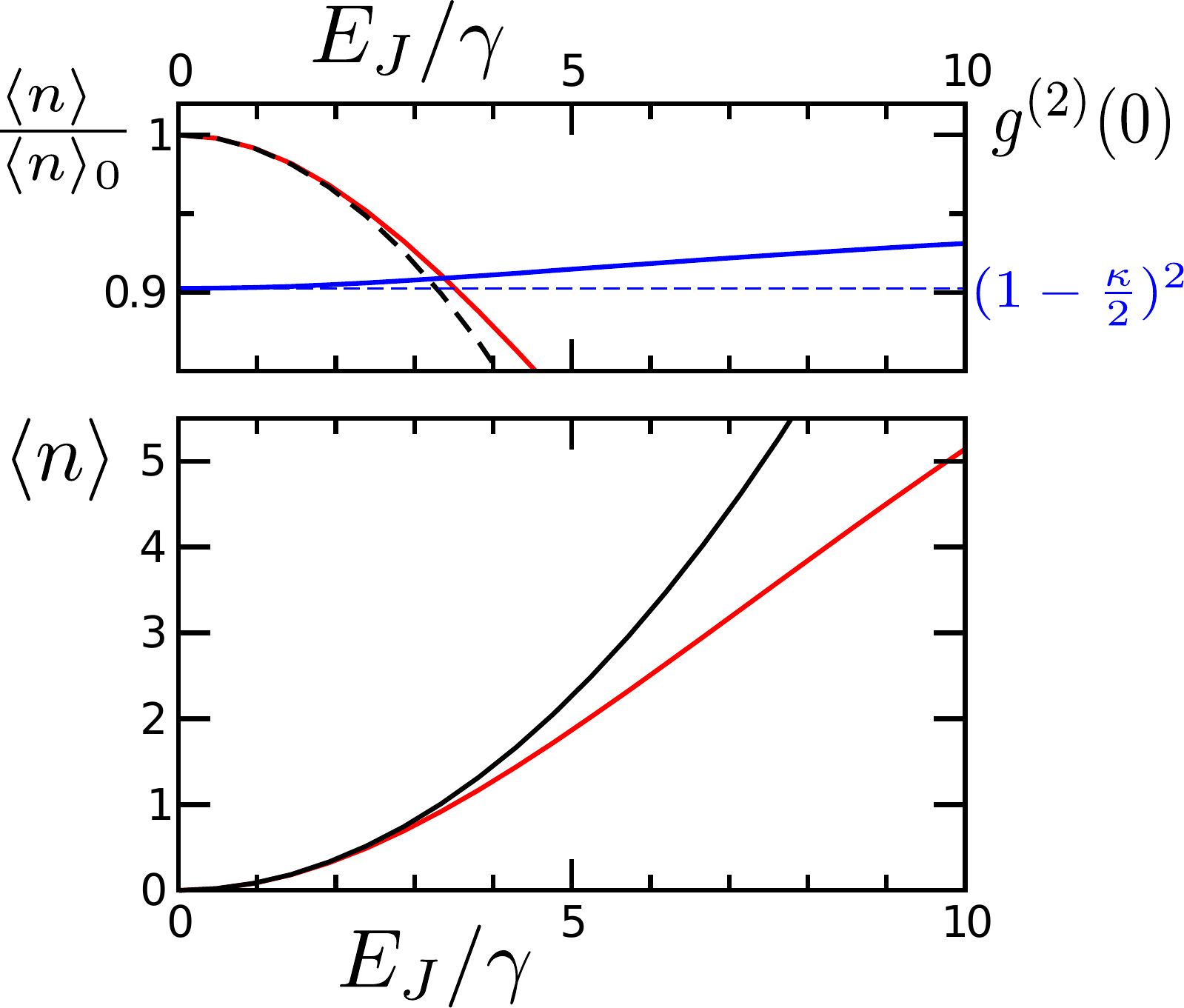}
\caption{\label{fig:Fig1}Occupation of a cavity with mode frequency $\omega_0$ and photon loss rate $\gamma$ driven by a JJ with Josephson energy $E_J$. Numerical data (red) are shown together with perturbative results in the CB-regime (black). Upper panel: Photon occupation normalized to the lowest order CB-result (\ref{n0}) together with the first correction (\ref{2eta2adagger}) (black, dashed). On the same scale, we display $g^{(2)}(0)$ for $\kappa=0.1$ (blue) together with the CB-prediction (\ref{eq:g2_1ph}) (blue, dashed) indicated on the right axis. Lower panel:  Photon occupation vs.\ driving strength together with the CB-prediction (\ref{n0}) (black).}
\end{figure}

Accordingly, an expansion of the Bessel functions in the above expressions allows for analytical results when only lowest and next lowest order terms  are taken into account. This implies
\begin{equation}
H_{1, CB}=\hbar\Delta \, n + i\frac{\tilde{E}_J}{2} \left[ : \left({\rm e}^{i\eta} a^\dagger -{\rm e}^{-i\eta} a\right)\, (1-n\frac{\kappa}{2}):\right]\,
\end{equation}
with $\tilde{E}_J=E_J^* \sqrt{\kappa}$. Note that by neglecting the $\kappa$-dependent correction and replacing
the ${\rm e}^{i\eta}$ operator by a c-number, this Hamiltonian would describe a driven harmonic oscillator in the RWA limit. The stationary state of such a driven, damped harmonic oscillator is a coherent state with Poissonian occupation and $g^{(2)}(0)\equiv1$. Further, one  has
\begin{eqnarray}\label{n2}
\langle n\rangle &=& \frac{\tilde{E}_J}{2\hbar\gamma} \langle :\left({\rm e}^{i\eta} a^\dagger +{\rm e}^{-i\eta} a\right) (1-n\frac{\kappa}{2}):\rangle\nonumber\\
\langle n^2\rangle -\langle n\rangle &=& \frac{\tilde{E}_J}{2\hbar\gamma}\langle :\left({\rm e}^{i\eta} a^\dagger +{\rm e}^{-i\eta} a\right) n:\rangle\, \left(1-\frac{\kappa}{2}\right)\, .
\end{eqnarray}

Based on the expansions above
, an analytical calculation of $g^{(2)}(0)$ is possible. It turns out that this requires knowledge of the moments $\langle {\rm e}^{i\eta} a^\dagger n\rangle\ , \ \langle {\rm e}^{2i\eta} (a^\dagger)^2\rangle$, and $ \langle {\rm e}^{i\eta} a^\dagger\rangle$, where the latter one appears also  in $\langle n\rangle$. In this order of a perturbative treatment, one only keeps terms up to $\tilde{E}_J^4$ and thus drops all contributions of the form $(a^\dagger)^n a^m$ with $n, m>2$.
\begin{figure}
\centering
\includegraphics[width=0.5\linewidth]{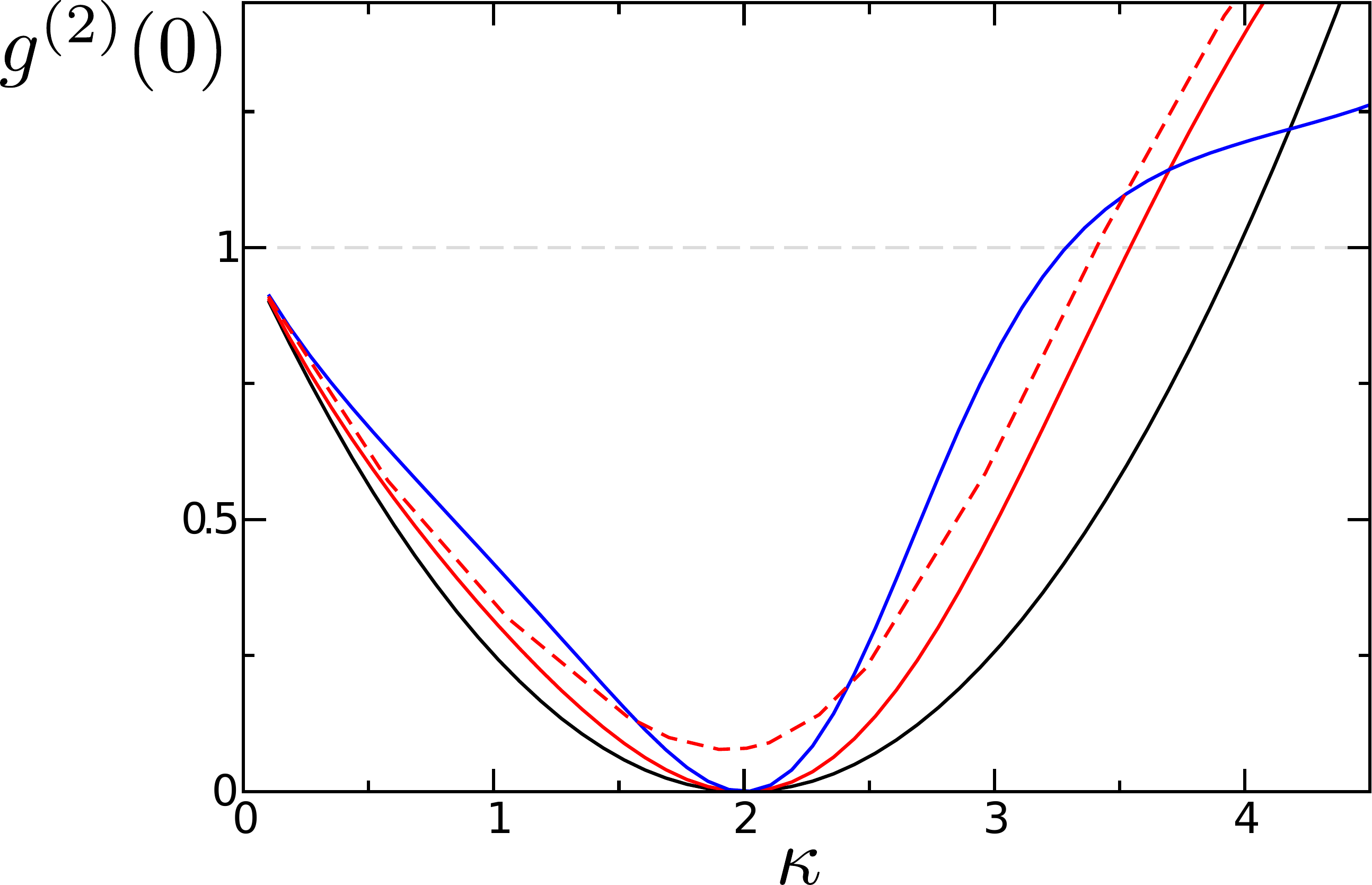}
\caption{\label{fig:Fig2a}Photon correlation  $g^{(2)}(0)$ vs.\ cavity ground state width $\kappa$ at the one-photon resonance for various driving strengths $E_J/\gamma$ (in terms of $\langle n \rangle_0$): CB-result (\ref{eq:g2_1ph}) for $\langle n \rangle_0=0.01$ (black) and numerical data (RWA) for $\langle n \rangle_0=0.5$ (red), $\langle n \rangle_0=2$ (blue). Full data (non-RWA) including also the impact of two-photon processes are depicted in red-dashed for  $\langle n \rangle_0=0.5, Q=10$.}
\end{figure}

This way, one obtains
\begin{eqnarray}
\langle {\rm e}^{i\eta} a^\dagger n\rangle &\approx &  \frac{\tilde{E}_J}{2\hbar} \frac{i}{\Delta+i\frac{3}{2}\gamma}\left[ \langle {\rm e}^{2i\eta} (a^\dagger)^2\rangle+2\langle n\rangle (1-\frac{\kappa}{2})\right]\, \nonumber\\
\langle {\rm e}^{2i\eta} (a^\dagger)^2\rangle &\approx &  \frac{\tilde{E}_J}{2\hbar} \frac{i}{\Delta+i\frac{1}{2}\gamma} \langle {\rm e}^{i\eta} a^\dagger\rangle (1-\frac{\kappa}{2})\, \nonumber\\
\langle n\rangle &\approx & \langle n\rangle_0\, \left[1-\kappa \langle n\rangle_0\right] \label{2eta2adagger}
\end{eqnarray}
with the lowest order CB-result
\begin{eqnarray}\label{n0}
\langle n\rangle_0 &=& \frac{\tilde{E}_J}{\hbar\gamma} {\rm Re}\{\langle {\rm e}^{i\eta} a^\dagger\rangle_0\}\nonumber\\
&=&\left(\frac{\tilde{E}_J}{\hbar\gamma}\right)^2 \frac{\gamma^2}{4 \Delta^2+\gamma^2}\, .
\end{eqnarray}
These expressions combine to yield at resonance $\Delta=0$ for (\ref{n2})
\begin{equation}
\langle n^2\rangle -\langle n\rangle = \langle n\rangle_0^2 \, \left(1-\frac{\kappa}{2}\right)^2\,
\end{equation}
which is of order $\tilde{E}_J^4$ and implies
\begin{equation}\label{eq:g2_1ph}
g^{(2)}(0) = \left(1-\frac{\kappa}{2}\right)^2\, .
\end{equation}
 Here, next order corrections are of order $\tilde{E}_J^2$. As displayed in Fig.~\ref{fig:Fig1}, these analytical predictions capture the exact data very accurately for low Josephson coupling or, equivalently, low driving of the cavity. The photon occupancy is suppressed, however, for stronger driving due to the anharmonicity of the Bessel function $J_1(2\sqrt{\kappa n})$ in (\ref{photon1}). Non-classical photonic states, i.e.\ $g^{(2)}(0)<1$, are produced in the cavity for $0<\kappa<4$, where at $\kappa=2$ one has complete anti-bunching even for strong driving (cf.~Fig.~\ref{fig:Fig2a}). In this situation the harmonic cavity effectively reduces to a two level system since the transition matrix element from the first to the second excited oscillator state vanishes exactly. Interestingly, when the full dynamics (non RWA) according to (\ref{exact}) is considered, contributions from off-resonant two photon processes induce at $\kappa=2$ a small finite $g^{(2)}(0)$.
 Note, that some of the weak-driving results above can alternatively be derived by directly employing a picture of excitation and decay rates . For instance, Eq.~(\ref{eq:g2_1ph}) then follows as,
 \[
 g^{(2)}(0)\approx \frac{2P_2}{P_1^2}\approx\frac{1}{2} \left|\frac{ \langle 2 |:a^\dagger\, (1-n\frac{\kappa}{2}):|1\rangle}{ \langle 1 |:a^\dagger\, (1-n\frac{\kappa}{2}):|0\rangle}\right|^2\;,
 \]
 where $P_n$ indicates the probability to find the cavity state $|n\rangle$ occupied.

\section{Two-photon resonance}

We now turn to the two-photon resonance $\omega_J=2 \omega_0$.  Following the procedure outlined above,  ones derives the RWA Hamiltonian
\begin{equation}
H_{2}=\hbar\tilde{\Delta} \, n + \frac{E_J^*}{2} \left[ : \left({\rm e}^{i\eta} (a^\dagger)^2 +{\rm e}^{-i\eta} a^2\right)\, \frac{J_2(2\sqrt{\kappa n})}{n}:\right]\,
\end{equation}
with de-tuning $\tilde{\Delta}=\omega_0-\omega_J/2$. In the regime of low photon occupancy (CB-regime), the above expression simplifies in leading order to
\begin{equation}
H_{2, CB}=\hbar\tilde{\Delta} \, n + \frac{\hat{E}_J}{4} \left[{\rm e}^{i\eta} (a^\dagger)^2 +{\rm e}^{-i\eta} a^2\right]\, ,
\end{equation}
where now $\hat{E}_J=E_J^*\, \kappa$. This Hamiltonian is, again with ${\rm e}^{i\eta}$ replaced by a c-number, identical to that of a parametrically driven oscillator (RWA-limit) which is well-known to induce squeezing (see below).
\begin{figure}
\centering
\includegraphics[width=0.5\linewidth]{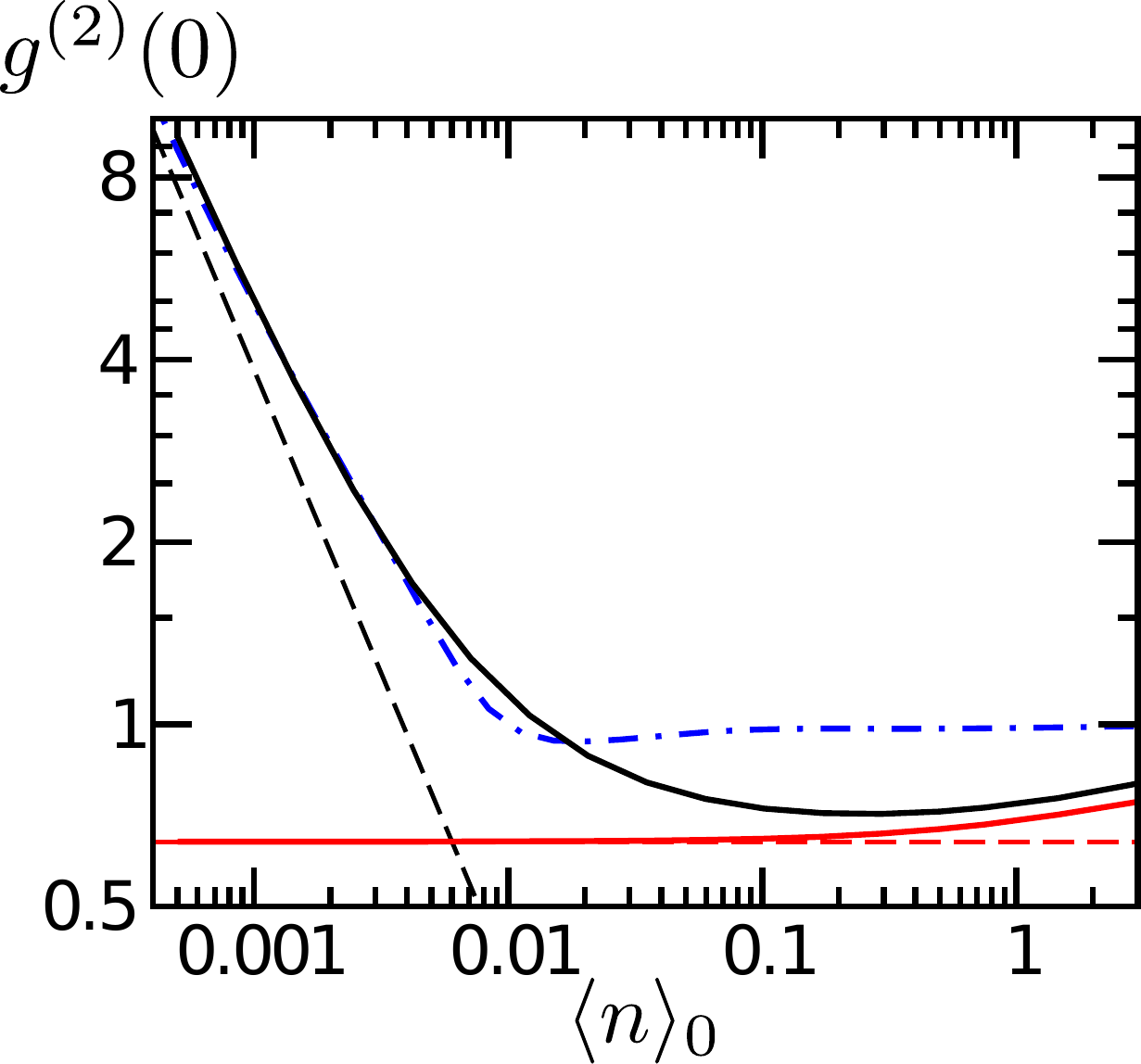}
\caption{\label{fig:Fig2b}
Photon correlation at the one-photon resonance vs. the Josephson coupling in terms of $\langle n\rangle_0$ [see (\ref{n0})] for $\kappa=0.4$ together with the result at the two-photon resonance (blue, dashed-dotted). The one-photon contribution (red, solid) reduces to the CB-result (\ref{eq:g2_1ph}) (red, dashed) for low driving. The full $g^{(2)}(0)$ (non-RWA) at the single-photon resonance $\omega_0=\omega_J$ (black, solid for $Q=5$) diverges for weak driving  due to the dominating impact of off-resonant two-photon processes in accordance with the perturbative result (\ref{g21ph}) (black dashed). For comparison, data at the two-photon resonance are obtained with a renormalized  $E_J \rightarrow E_J \,\sqrt{1+Q^2}/\kappa$.}
\end{figure}

Analytical results based on $H_{2, CB}$ are easily obtained. At the two-photon resonance, photon noise in the CB-regime is already dominated by the leading order contributions in an expansion of the Bessel functions and we gain at resonance $\tilde{\Delta}=0$
\begin{eqnarray}
\langle n\rangle_0 &=& -i \frac{\hat{E}_J}{2\hbar\gamma} \langle {\rm e}^{i\eta} (a^\dagger)^2 -{\rm e}^{-i\eta} a^2\rangle\nonumber\\
&=&\frac{1}{2}\left(\frac{\hat{E}_J}{\hbar\gamma}\right)^2\,
\end{eqnarray}
and
\begin{eqnarray}\label{n22ph}
\langle n^2\rangle &=& i \frac{\hat{E}_J}{2\hbar\gamma} \langle {\rm e}^{i\eta} (a^\dagger)^2 +(a^\dagger)^2 n +
{\rm e}^{-i\eta} ( a^2+n a^2)\rangle+\frac{1}{2}\langle n\rangle_0\nonumber\\
&\approx & \frac{3}{2}\langle n\rangle_0  + i \frac{\hat{E}_J}{2\hbar\gamma} \langle {\rm e}^{i\eta} (a^\dagger)^2 n +
{\rm e}^{-i\eta}n a^2\rangle\, .
\end{eqnarray}
Since the last term is of order $\hat{E}_J^4$, this provides
\begin{equation}\label{g22ph}
g^{(2)}(0) = \frac{1}{2\langle n\rangle_0}\,
\end{equation}
being of order $1/\hat{E}_J^2$ with corrections of order 1. Hence, in the weak driving regime the photon correlation at the two-photon resonance diverges and radiated light becomes strongly bunched (see Fig.~\ref{fig:Fig2b}).
 In fact, for weak driving the numerator of $g^{(2)}(0)=\langle n(n-1)\rangle/\langle n\rangle^2$, given by twice the two-photon occupation probability, $2P_2$, is strongly enhanced by two-photon creation processes. These processes will completely dominate photon correlations even around the single-photon resonance according to numerical non-RWA results.
 The full picture for the photon-correlation function  $g^{(2)}(0)$ at the one-photon resonance, see Fig.~\ref{fig:Fig2b}, is then the following: The RWA result for the one-photon process (red, solid line in Fig.~\ref{fig:Fig2b}), which reduces  to the CB-result (\ref{eq:g2_1ph}) (red, dashed) for low driving only holds for {\em high quality} cavities. For cavities with only moderate $Q$ values (cf. black, solid line for $Q=5$), for weak driving the numerator of  $g^{(2)}(0)$  becomes determined by the (off-resonant) contribution of the two-photon process [cf. Eq.~(\ref{n22ph})], while its denominator is still well described by the one-photon result, Eq.~(\ref{n0}). This leads to a crossover (black-dashed) to
 \begin{equation}\label{g21ph}
g^{(2)}(0) \approx \frac{\kappa}{4\langle n\rangle_0} \frac{1}{1+Q^2}\,,
\end{equation}
 The similarity to the $g^{(2)}(0)$ divergence described by Eq.~(\ref{g22ph}) is demonstrated by comparison to a (suitably renormalized) result at the two-photon resonance (blue, dashed-dotted).

\section{Photon number distribution}\label{sec3}

\begin{figure}[ht]
\centering
\includegraphics[width=\linewidth]{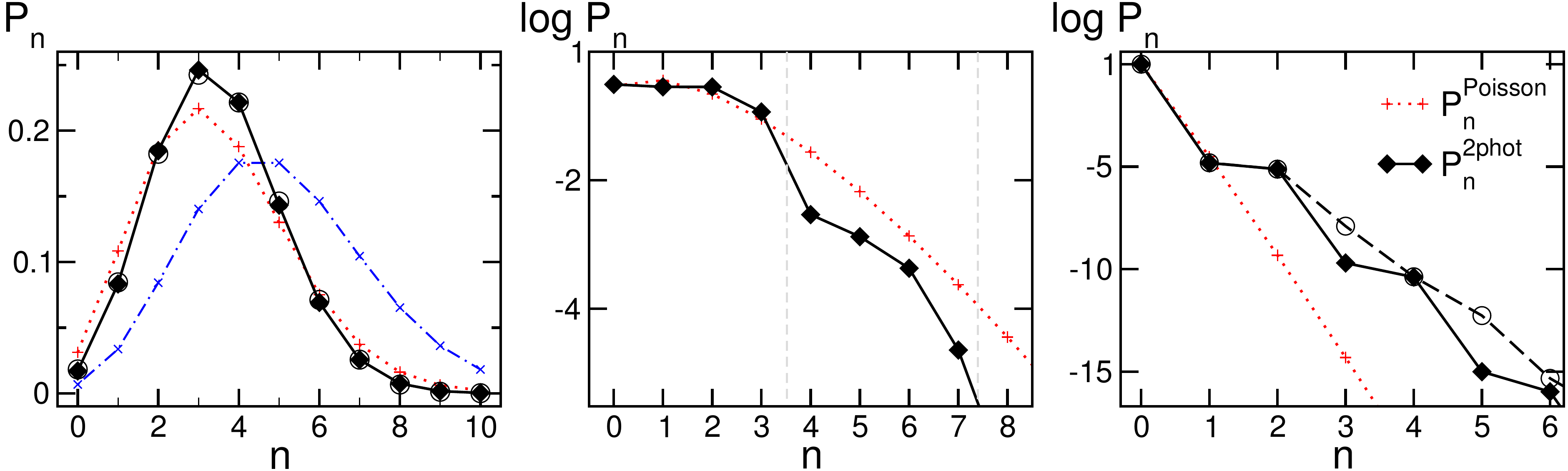}
\caption{\label{fig:Fig3}Photon number distribution $P_n$ at the single-photon (left and middle panel) and the two-photon resonance (right panel). Left: $P_n$ for  $\langle n \rangle_0=5 $ and $\kappa=0.1$ (black diamonds). In comparison, Poissonian distributions with same $\langle n \rangle$ (red crosses) and  with same $\langle n \rangle_0$ (blue crosses) together with the distribution of an amplitude squeezed state (two-photon coherent state) with same $\langle n \rangle$ and $g^{(2)}(0)=0.92$ (black circles), see text for details. Middle: $P_n$ for  $\kappa=3.5$ and strong driving $\langle n \rangle_0=50$ (black diamonds) in comparison to the Poissonian occupation of a linear drive (red-dashed). Zeros of $J_1(2\sqrt{\kappa n})$ at certain values $n=n_k, k=1,2,\ldots$ lead to a suppression of the occupation with a dip at $n_1=1$ and pronounced steps at $n_2= 3.5$ and $n_3= 7.4$ (grey). Right: $P_n$ at the two-photon resonance $\omega_J=2\omega_0$ showing a pronounced even-odd effect within an RWA  (diamonds) and a full treatment (circles) for  $Q=10$.}
\end{figure}

To obtain a deeper insight into the non-classical nature of the cavity photons, we now analyze in more detail the corresponding occupation number distribution $P_n$ at the single and the two-photon resonances, respectively (see Fig.~\ref{fig:Fig3}).

In the first case, for moderate driving and ground state widths of the cavity, the combined dynamics of cavity+JJ creates amplitude squeezed states (two-photon coherent states) \cite{milburn:1994}. Formally, these states result from the vacuum state via $S(r)\, D(\xi)|0\rangle$ by creating first a coherent state with $D(\xi)= \exp(\xi a^\dagger-\xi^* a)$ and then a squeezed state with $S(r)=\exp[\frac{1}{2} (r^* a^2 - r a^{\dagger\, 2})]$. Here, from the numerical data one extracts real values for $\xi, r$ with: $\xi\approx 1.859, r\approx 0.162$. The corresponding photon distribution is non-Poissonian with $g^{(2)}(0)<1$.

For strong driving and large $\kappa$, $P_n$ is determined by strong nonlinearities, where the sub-linear behavior of the Bessel-function in (\ref{photon1}) leads to a depletion of higher lying states compared to a linear drive. It even induces  pronounced downwards steps in $P_n$  approximately at those $n$-values where $J_1(2\sqrt{\kappa  n})=0$ so that certain transition elements between Fock states are strongly suppressed.

At the two-photon resonance, deviations from the Poissonian profile are more pronounced due to strong squeezing and a pronounced even-odd effect can be observed \cite{Gramich:2013}. In particular, the occupation of the second excited state is strongly enhanced in comparison to the Poissonian case. We have already seen above that this leads in the weak driving regime to diverging photon correlations (cf.~also \cite{Padurariu:2012,Leppakangas:2013,Leppakangas:2014}).

\section{Conclusions}
We studied the creation of light  by Cooper-pair tunneling across a voltage-biased Josephson junction coupled to a superconducting microwave resonator. Already in the  weakly driven limit, where Cooper pair transfer occurs sequentially, the nonlinearity of the Josephson junction can yield strongly correlated and non-classical light. For stronger driving, back-action from the cavity field on the Cooper pair tunneling becomes relevant leading to complex nonlinear dynamics of the coupled system. We derived RWA Hamiltonians to capture one and two-photon creation processes at the corresponding resonances and found explicit analytical results for the photon correlation functions in the weak driving limit. These show, that the device can create completely anti-bunched (non-classical) light, as well as strong bunching. The robustness of these features and the crossover and interplay of various creation processes were studied within a numerical approach. The full photon distribution in various regimes, reflect squeezing, the full nonlinearity of the Bessel function appearing in the RWA Hamiltonian (which is explored for stronger driving), and pronounced even-odd effects, respectively.
In addition to their inherent interest, the rich variety of non-classical states of light created in this device may be  observed and characterized to investigate the dynamics of the underlying charge transfer process.

\section*{Acknowledgements} Fruitful discussions with D. Esteve, M. Hofheinz, and F. Portier are gratefully acknowledged. Financial support was provided by the SFB/TRR-21 and the DFG under AN336/6-1 and AN336/7-1.

\section*{References}
\bibliographystyle{unsrt}
\bibliography{bib_proceedings}
\end{document}